\documentclass[letterpaper, 9pt, conference]{ieeeconf}  

\IEEEoverridecommandlockouts                              

\usepackage{gensymb}
\usepackage{enumerate}

\usepackage{graphicx}
\usepackage{subcaption}
\usepackage{color,soul}
\usepackage{dblfloatfix}

\usepackage{multirow}

\title{\Huge 
RF-Trojan: Leaking Kernel Data Using Register File Trojan}

\author{Mohammad Nasim Imtiaz Khan, Asmit De and Swaroop Ghosh\\
School of EECS, The Pennsylvania State University, Uiversity Park, PA.\\
Email: muk392@psu.edu, asmit@psu.edu and szg212@psu.edu}

\begin{document}
\maketitle
\thispagestyle{empty}
\pagestyle{empty}

\begin{abstract}
Register Files (RFs) are the most frequently accessed memories in a microprocessor for fast and efficient computation and control logic. Segment registers and control registers are especially critical for maintaining the CPU mode of execution that determines the access privileges. In this work, we explore the vulnerabilities in RF and propose a class of hardware Trojans which can inject faults during read or retention mode. The Trojan trigger is activated if one pre-selected address of L1 data-cache is hammered for certain number of times. The trigger evades post-silicon test since the required number of hammering to trigger is significantly high even under process and temperature variation. Once activated, the trigger can deliver payloads to cause Bitcell Corruption (BC) and inject read error by Read Port (RP) and Local Bitline (LBL). We model the Trojan in GEM5 architectural simulator performing a privilege escalation. We propose countermeasures such as read verification leveraging multi-port feature, securing control and segment registers by hashing and L1 address obfuscation.
\end{abstract}

\section{Introduction}
Hardware Trojan \cite{Trojan_war} is a malicious modification in a circuit that causes a chip to perform undesirable operations. Ideally, these modifications made to an Integrated Circuit (IC) should be detected during pre-Silicon verification and post-Silicon testing. In order to evade such structural and functional testing, an adversary designs the Trojan to activate only under certain rare conditions and to remain undetected during the test phase. For example, the analog Trojan trigger proposed in {\cite{A2_paper}} charges a capacitor every time an instruction is being executed. After few cycles, the capacitor charges up and asserts a signal which is used to flip some specific bits of control logic and can escalate the adversary's user privilege.

Hardware Trojan is composed of two parts: Trigger and Payload \cite{tehranipoor_wang_2012, Chakra_HT}. A Trojan trigger similar to \cite{DATE_Nasim_2019_blind_review} has been considered in this work (details in Section II.A). 
Once triggered, the Trojan delivers payloads to the Register File (RF) such as, Bitcell Corruption (BC), Read Port (RP) and Local Bitline (LBL) Trojans. The RP and LBL Trojans inject read errors. Note that we have considered the trigger proposed in \cite{DATE_Nasim_2019_blind_review} (over \cite{A2_paper}) since it, i)  is robust against process and temperature variation; ii) evades post silicon testing and system level detection mechanisms; and, iii) incurs less area overhead.

We note that RF stores security critical information and a tampering can lead to leakage of sensitive data. For example, a code segment (CS) register file contains a Current Privilege Level (CPL) field that determines whether the CPU is currently executing in user mode or kernel mode. User mode processes are restricted from accessing data from the kernel space based on the CPL set in the CS register. The adversary can take control of the kernel mode by manipulating the RF entry that stores the execution mode and run unauthorized operations. 

\textbf{Attack Model:} We have assumed that the Trojan trigger and payload has been either inserted by the designer or by the untrusted fabrication house. The adversary is a user who is sponsored by the fabrication house and is aware of the trigger requirements. After the deployment of the chip in the market, adversary can launch a malicious program to activate the trigger. The adversary can then deploy the desired payloads using the proposed BC/RP/LBL Trojans. Note that even if the trigger is activated, BC/RP/LBL Trojans can remain dormant (until payload deployment conditions are met) and the system functions normally. The Trojan payload changes the CPL field in the CS register from 3 (user mode) to 0 (kernel mode). This essentially escalates the privilege of adversary's process and allows access to kernel space.

Note that the work {\cite{A2_paper}} also proposes a Trojan which escalates user privilege. However, the differences are: 
i) the trigger {\cite{A2_paper}} works for a specific architecture where Supervisor Register (i.e. payload) to breach security is implemented using  D-flip flops, however, modern processors employ RF to store control bits that is considered in this work; ii) the trigger in {\cite{A2_paper}} is activated after a repeated sequence of unlikely events such as division-by-zero, whereas {\cite{DATE_Nasim_2019_blind_review}} hammers a L1 cache address with unique data pattern. The trigger in \cite{A2_paper} can be detected by a routine that counts these events to flag a security issue. We consider address hammering \cite{DATE_Nasim_2019_blind_review} which is extremely difficult to detect since keeping track of each address is nearly impossible. 

In summary, following contributions are made in this paper. We, (a) summarize the vulnerabilities of RF design which can be exploited to insert Trojan; (b) propose information corruption by BC Trojan and inject read error by RP and LBL Trojan; (c) demonstrate a privilege escalation attack with an exploit code in GEM5; (d) propose detection and countermeasures such as, hashing, validation using redundant read port and obfuscation of L1 addresses.


\begin{figure} [b] 
 \vspace{-4mm}
 \begin{center}
    \includegraphics[width=.48\textwidth]{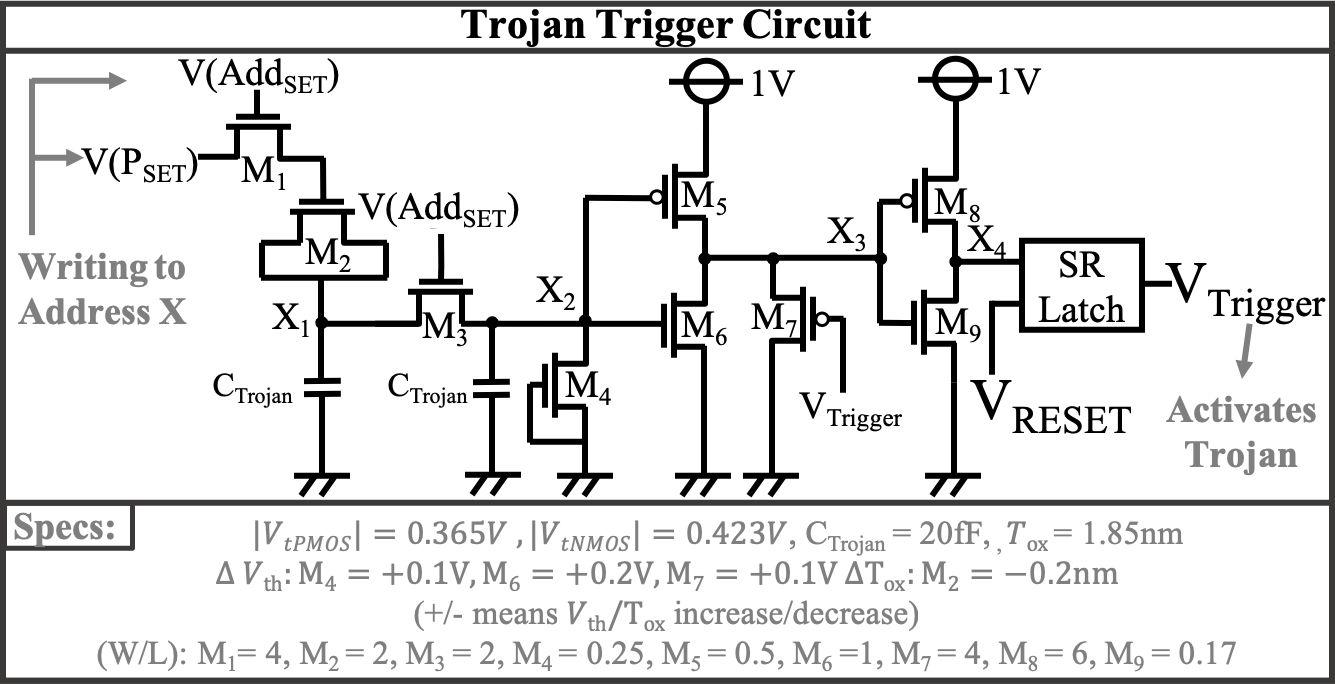}
 \end{center}
 \vspace{-4mm}
 \caption{Capacitor-based Trojan trigger circuit \cite{DATE_Nasim_2019_blind_review}.} 
 \label{Trigger_DATE}
  \vspace{-2mm}
 \end{figure}

The paper is organized as follows: Section II presents background on Trojan trigger and system architecture; Section III describes background on RF and explores its vulnerability; Section IV describes the proposed RF Trojans; Section V presents the attack demonstration using GEM5. Section VI presents a discussion on the practicality of the proposed Trojans and possible defenses; and finally, Section VII draws the conclusion.

\section{Background}

\subsection{Trigger Design \cite{DATE_Nasim_2019_blind_review}}
The trigger circuit (Fig. \ref{Trigger_DATE}) is designed to be activated if a particular memory address ($Add_{SET}$, chosen during Trojan design) is written with a specific data pattern, $P\textsubscript{SET}$ for at least $N_{SET}$ times. The trigger has two inputs namely, $V(Add\textsubscript{SET})$ and $V(P\textsubscript{SET})$. $V(Add\textsubscript{SET})$ (= 1V) is the wordline enable signal of $Add_{SET}$. For $V(P\textsubscript{SET})$, a simple logic circuit can be implemented which outputs logic 1 (1V) if a specific data pattern is sent to data bus. For example, let's consider that the data bus width is 8-bits. Assume that four specific data bits are taken to design the trigger logic e.g., data[0], data[3], data[4] and data[6]. The logic circuit will output `1' if data[0] = data[3] = data[6] = 1 and data[4] = 0. In practice, data bits with low activation probabilities can be used to design the logic to lower the overall probability of assertion unless intended.

Whenever $Add_{SET}$ is written with $P\textsubscript{SET}$ data pattern, MOSFETs $M_{1}$ and $M_{3}$ turn ON and $C\textsubscript{Trojan}$ charges up from the $V(P\textsubscript{SET})$
via Fowler Nordheim (FN) tunneling \cite{Ravindra_tunneling_relaxed} through $M_{2}$. Note that $M_{2}$ has a thinner gate oxide compared to other MOSFETs and its source and drain are shorted (i.e. $M_2$ behaves as a capacitor). $M_{4}$ is an OFF transistor which offsets the gate leakage of $M_{5}$ and prevents unwanted charging-up of node $X_{2}$. $M_{7}$ keeps node $X_3$ as low as possible until node $X_{2}$ charges up sufficiently. The node $X_4$, that is charged up during the hammering process, is used as the SET input of an SR latch. SR latch output ($V_{Trigger}$) transitions from $0\rightarrow1$ when $X_4$ charges up to $\sim$0.5V which requires $N_{SET} = 1837$. The signal $V_{Trigger}$ is then used to deploy the RF Trojan payloads. 

The work \cite{DATE_Nasim_2019_blind_review} shows that adversary hammers for a duration of $T\textsubscript{ON}$ and then stays idle for $T\textsubscript{OFF}$ and repeats this cycle. However, the circuit can still be triggered (minimum $T\textsubscript{ON}$\% = 30) but with a higher $N_{SET}$ since $C\textsubscript{Trojan}$ does not significantly leak during the OFF cycle.  Therefore, it becomes even harder to prevent this Trojan activation using system level techniques such as limiting consecutive accesses to one particular address up to a threshold. 

$V_{RESET}$ can be generated to deactivate the trigger by writing to $Add_{RESET}$ for at least $N_{RESET}$ times with a specific data pattern, $P\textsubscript{RESET}$ and using a circuit similar to the trigger one. A smaller $C_{Trojan}$ ($\sim$1fF) can be used in the RESET circuit to minimize the area ($N\textsubscript{RESET} = $ 92). However, the AND'ed output of $V(Add_{RESET})$ and $V(P_{RESET})$ can also serve as $V_{RESET}$.  

The trigger is robust and evades testing under worst-case process and temperature variation {\cite{DATE_Nasim_2019_blind_review}}. Trigger area/static power are both $<0.0001\%$ of that of a typical memory. Therefore, it has been noted that the trigger can pass optical inspection/side channel analysis.

\subsection{System Architecture Overview}
We consider a standard X86 microprocessor architecture with with L1 and L2 caches, where the L1 cache is further separated into L1 instruction cache and L1 data cache. The L1 cache is virtually indexed and physically tagged to improve performance. The system runs a standard Linux kernel with paging enabled and a flat linear address space (transparent segmentation). The memory management unit (MMU) in the microprocessor has a paging unit with fully associative TLB to speed up address translation, and memory protection is enabled in the paging unit.

\begin{figure}
       \vspace{0mm}
        \begin{subfigure}[b]{0.49\textwidth}
                (a) \centering
                \includegraphics[width=0.9\textwidth]{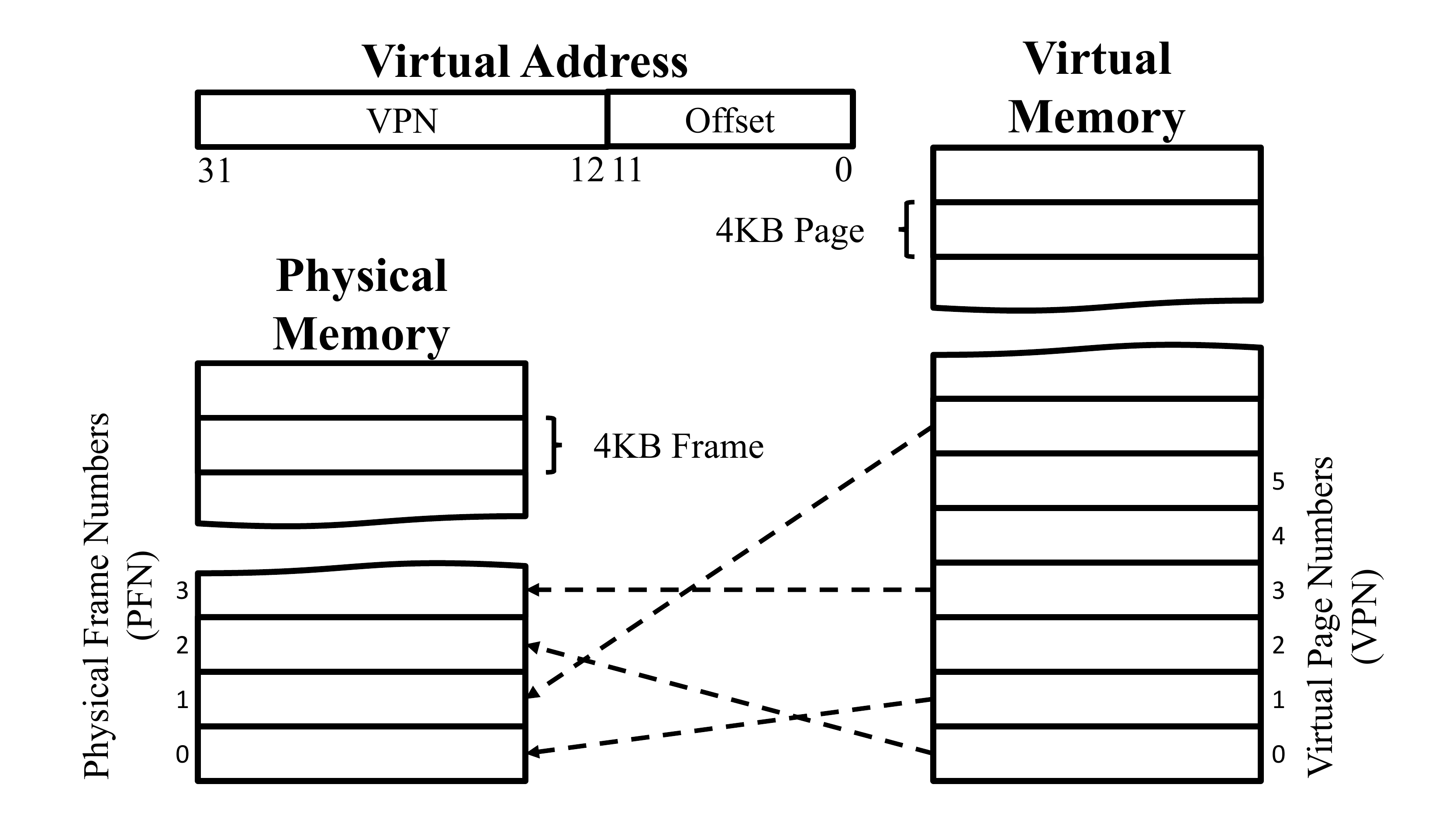}
        \end{subfigure}\\ \vspace{0mm}
        \hspace{-1mm}
        \begin{subfigure}[b]{0.49\textwidth}
                 (b) \centering
                \includegraphics[trim=0 5.625in 0 0,clip,width=0.9\textwidth]{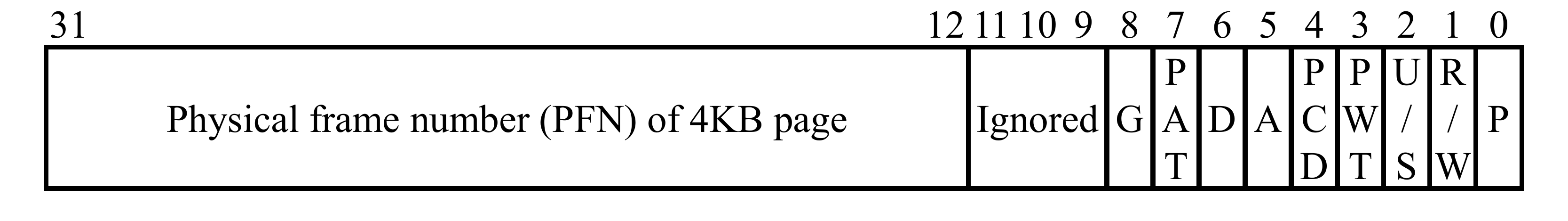}
        \end{subfigure}
        \vspace{-2mm}
        \caption{(a) Paging in a 32-bit address space with 4KB pages, and (b) Page Table Entry (PTE) fields for Intel X86 32-bit paging. \iffalse and, (c) an example TLB entry configuration showing VPN, PFN and the associated metadata bits. \fi}
        \label{paging}
        \vspace{-6mm}
\end{figure}

\textbf{Memory Layout:} Applications running on a specific CPU architecture have their complete view of the memory space (known as the virtual address space) addressable by the CPU. 
The installed physical memory (and the physical address space) is limited, and so the virtual address space is divided into smaller chunks known as virtual pages (Fig. \ref{paging}a) that are 
mapped to the physical memory as physical frames. The corresponding mapping information (virtual page number $\,\to\,$ physical frame number) is stored in a per-process page table in the OS kernel. The page table entries (PTE) also store some associated metadata bits signifying protection status, caching information, etc. as shown in Fig. \ref{paging}b. When the CPU accesses a memory address, the corresponding page mapping is pulled from the page table in the kernel and cached onto the TLB in the CPU.
In a 4GB virtual address space, the higher order 1GB of addresses are used by the kernel and is known as the kernel space. The remaining 3GB are given to the user processes and is known as the user space. The user space is segmented into several segments which may contain readable or writable data, or executable code. The kernel space also has its own code and data segments. Note that user space code has access to the data in the user space only, and no access to the data in the kernel space. However, kernel space code can access data from both the kernel and user space.

\textbf{CPU Protection Rings:} The operating system and the CPU work in conjunction to decide what user mode applications have access to. This protection is maintained in terms of privilege levels, that protects three primary resources: memory, I/O, and permission to execute certain machine instructions. In the x86 architecture, there are four privilege levels or protection rings: 0 (highest privilege) to 3 (lowest privilege); although, modern x86 OS kernels use only rings 0 (for kernel mode) and 3 (for user mode). At any time, the CPU is configured to run in either of these two rings, which determines the Current Privilege Level (CPL) of the code in execution. The CPU keeps track of the CPL by using segment descriptors stored in segment registers. The code segment (CS) register has a 2-bit field that specifies the CPL. When the kernel (user) code is executing, the CS register is updated and the CPL is set to 0 (3). 

\textbf{Memory Protection with Paging:} Memory protection in modern systems is primarily handled at the paging unit which is responsible for converting a linear virtual address to a physical address.
Bits 2 (U/S) and 1 (R/W) in the PTE of each page signify the protection status. The U/S bit 
is set (1) for kernel pages and unset (0) for user pages. The R/W  bit, although not responsible for setting privileges, is used to mark executable code pages as non-writable, so that a process cannot overwrite existing code. For RF-Trojan design, we are primarily concerned with the U/S bit in the PTE. When an address belonging to a particular page is accessed by some code, the corresponding PTE is read from the kernel and cached onto the TLB. The paging unit compares the CPL of the code set in the CS register with the U/S bit set in the PTE. If the U/S bit is set to 1, access to the data page is given to the code only if CPL is set to 0. This ensures that a segmentation fault is caused if the user code (with CPL 3) accesses kernel data pages (with U/S bit = 1).


\section{Register File: Basics and Vulnerabilities}
In this section, we present RF basics and its architecture and point out its vulnerabilities that can be exploited to implement Trojans.  

\subsection{Basics of Register File}
RF is fastest and most frequently accessed memory in the memory hierarchy where CPU holds temporary data and addresses. In every clock cycle, CPU may need to read multiple registers (for arithmetic instructions) to cater to the instructions and concurrently, write the outputs from previous instructions to the register file. Therefore, each register requires multiple read/write ports so that multiple entries can be read 
simultaneously by different instructions. 
The RF is designed in a way that only one write port can write at a time and write data can be bypassed to read ports using bypass-mux while writing if a processor needs to read that register in the same clock cycle. Modern superscalar processors require  more than 6 ports in the register file. RF is implemented using two back to back inverters similar to SRAM. However, multiple read/write ports are employed as mentioned before. The local and global bitlines employ domino logic for faster operation \cite{RAM_RF_Main}.

\begin{figure} [t] 
\vspace{-1mm}
\begin{center}
\includegraphics[width=.45\textwidth]{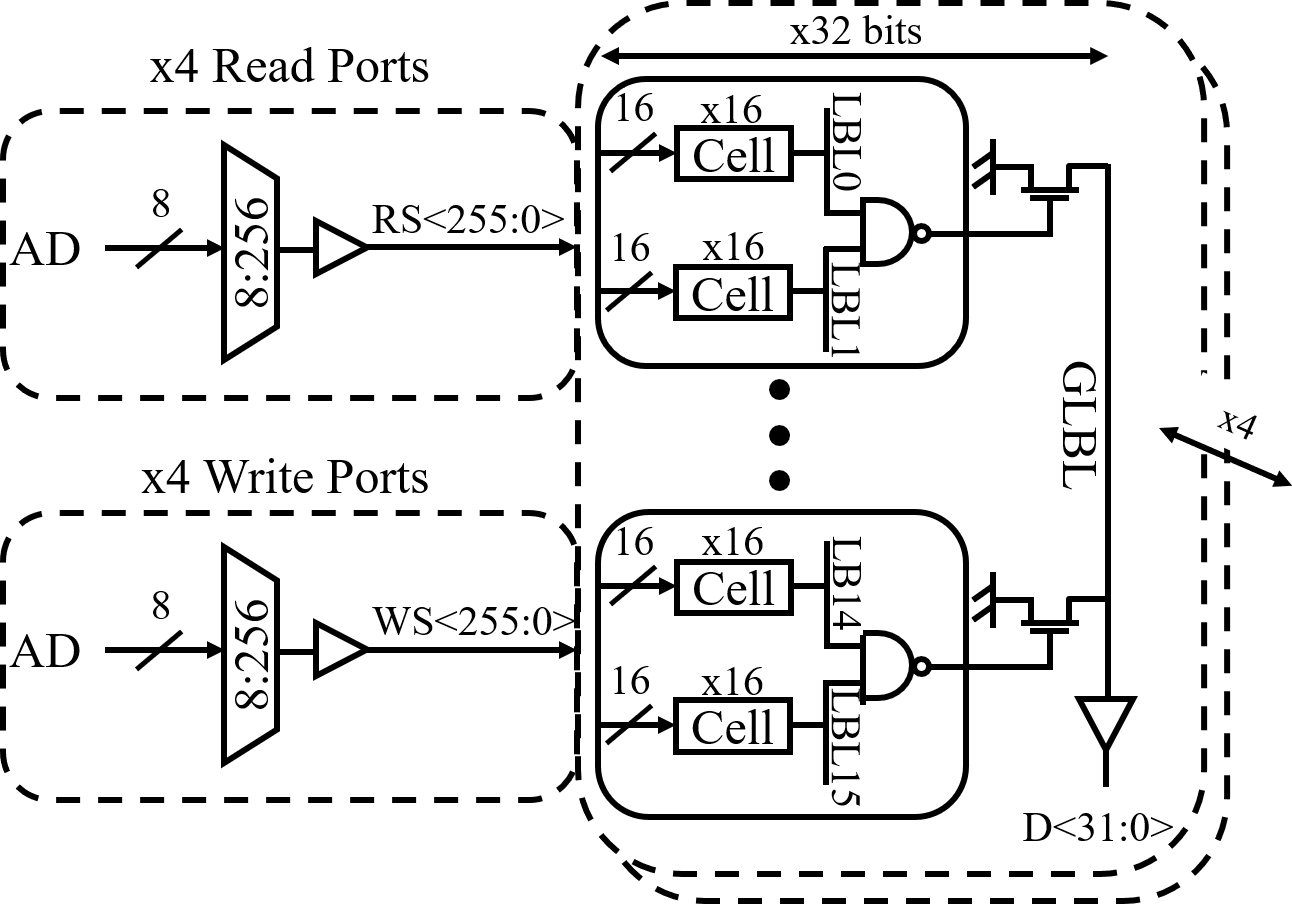}
\end{center}
\vspace{-3mm}
\caption{Register file architecture with 4 read and 4 write ports \cite{RAM_RF_Main}.} \label{rf_whole}
\vspace{-1.3mm}
\end{figure}

\begin{figure}[t]
        \centering
        \hspace{-1mm}
        \begin{subfigure}[b]{0.49\linewidth}
                \centering
                \includegraphics[width=0.99\linewidth]{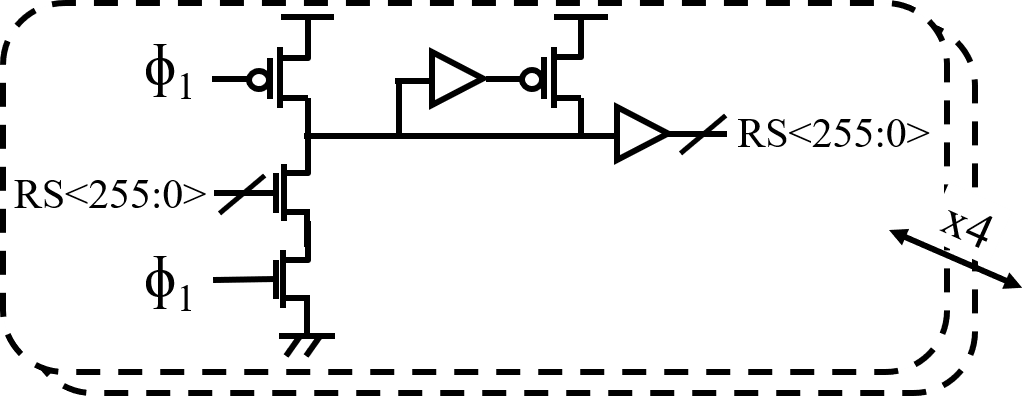}
                \caption{}
                \label{read_select}
        \end{subfigure}%
        \hspace{-1mm}
        \begin{subfigure}[b]{0.49\linewidth}
                \centering
                \includegraphics[width=0.99\linewidth]{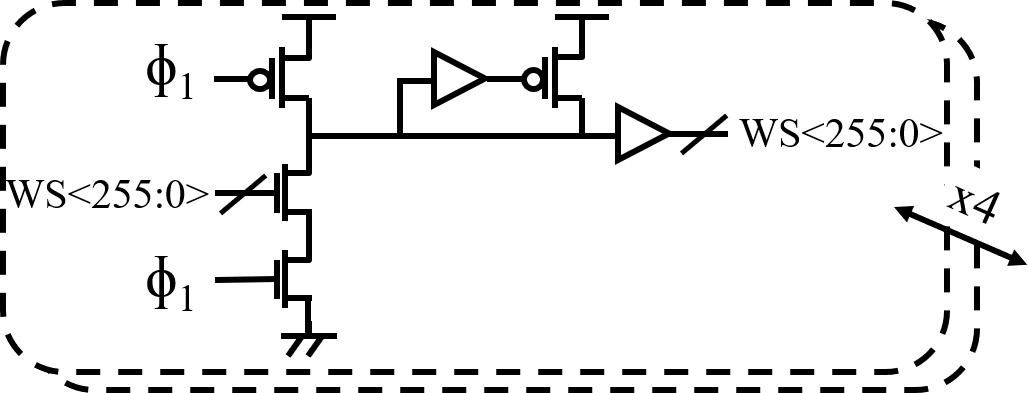}
                \caption{}
                \label{write_select}
        \end{subfigure}\\
        \vspace{-2mm}
        \caption{(a) RS  and (b) WS generation for reading/writing an address.}
	\vspace{-5mm}
        \label{rf_read_write_select}
\end{figure}

\begin{figure}[t]
        \centering
        \hspace{-1mm}
        \begin{subfigure}[b]{0.3\linewidth}
                \centering
                \includegraphics[width=0.99\linewidth]{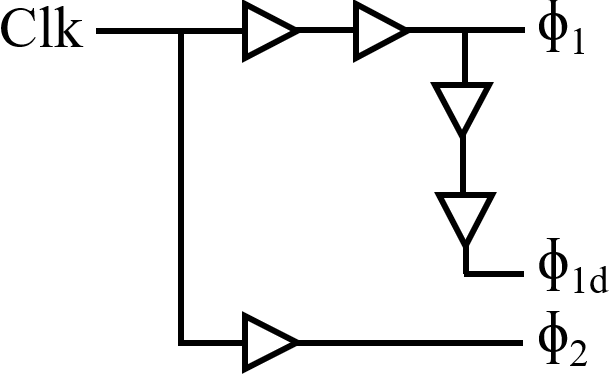}
                \caption{}
                \label{phase_signal}
        \end{subfigure}%
        \hspace{-1mm}
        \begin{subfigure}[b]{0.68\linewidth}
                \centering
                \includegraphics[width=0.99\linewidth]{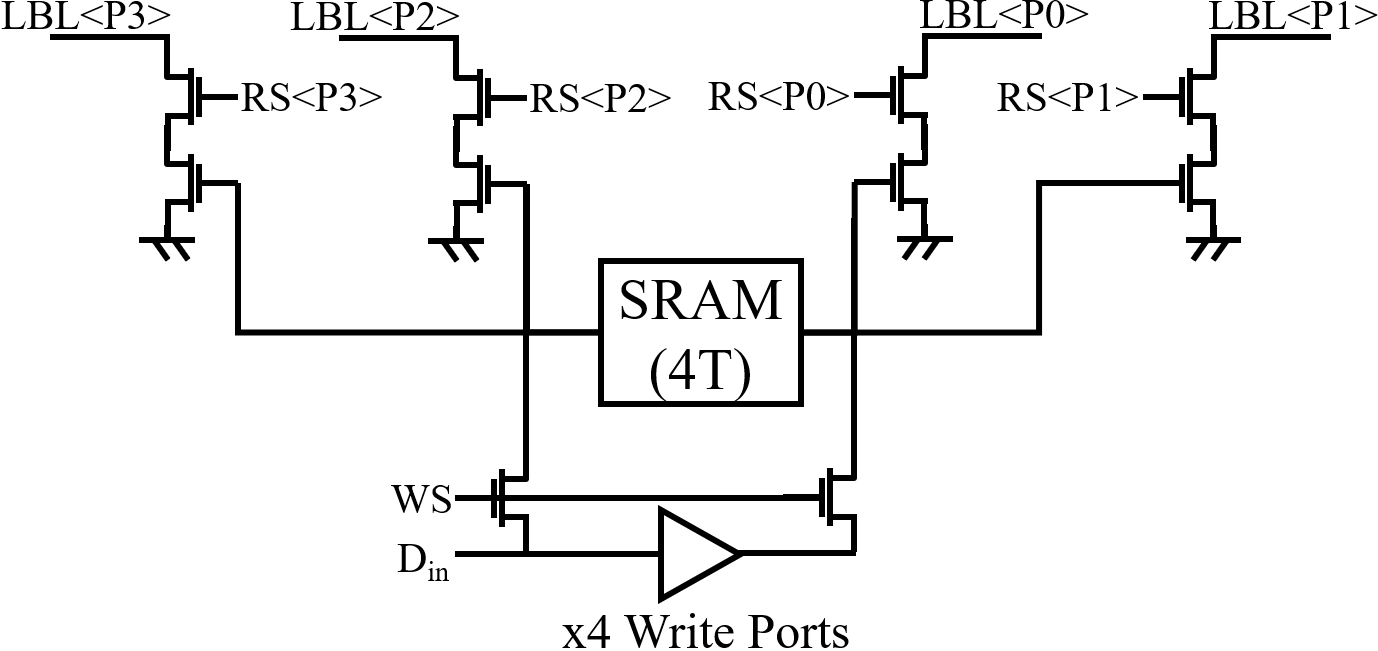}
                \caption{}
                \label{read_ports_4}
        \end{subfigure}\\
        \vspace{-2mm}
        \caption{(a) Phase signal generation for read/write operation; (b) 4 read ports for each bitcell.}
	\vspace{-3mm}
        \label{phase_read_ports_4}
\end{figure}

\begin{figure}[t]
        \centering
        \hspace{-1mm}
        \begin{subfigure}[b]{0.61\linewidth}
                \centering
                \includegraphics[width=0.99\linewidth]{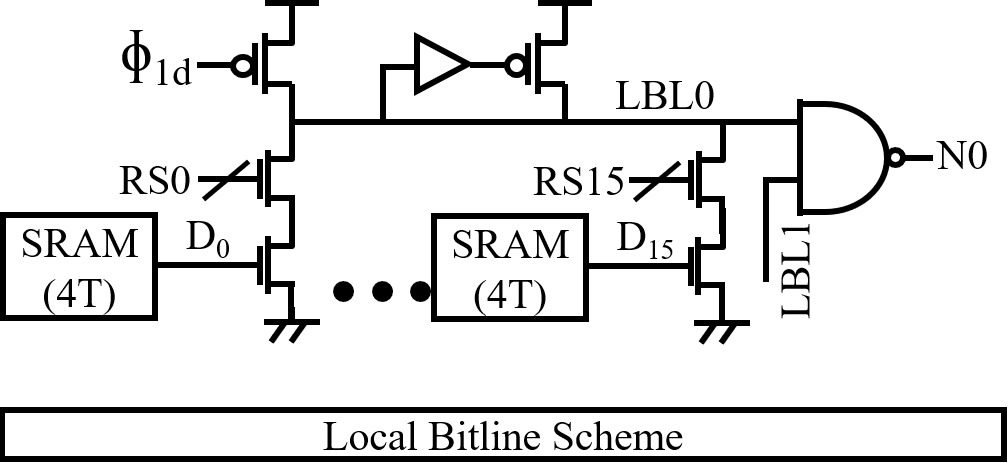}
                \caption{}
                \label{Local_bitline_sceme}
        \end{subfigure}%
        \hspace{-1mm}
        \begin{subfigure}[b]{0.36\linewidth}
                \centering
                \includegraphics[width=0.99\linewidth]{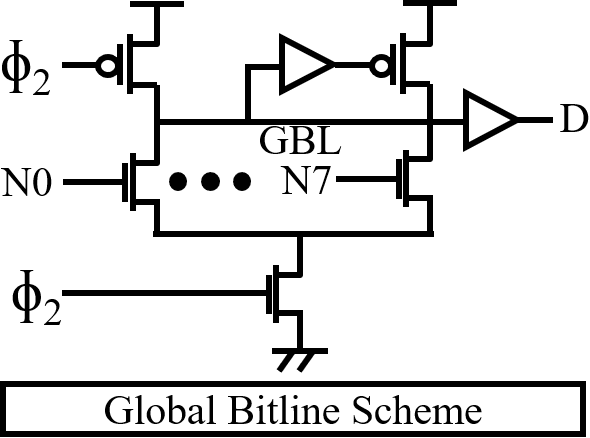}
                \caption{}
                \label{global_bitline_scheme}
        \end{subfigure}\\
        \vspace{-2mm}
        \caption{(a) LBL scheme, and (b) GBL scheme for read operation.}
	\vspace{-6mm}
        \label{local_global}
\end{figure}

\subsection{RF Architecture}
In this work, we have implemented 256 $\times$ 32-bits/word RF with 4 read and 4 write ports  \cite{RAM_RF_Main} using 22nm PTM technology \cite{22nm} in HSPICE (Fig. \ref{rf_whole}). The 8:256 address decoders per port is leveraged to send the Read/Write Select (RS/WS) signals to the array. Each Local Bitline (LBL) (one per read port) is connected with 16 register cells. There are 16 such local bitlines for each Global Bitline (GBL) making 16 $\times$ 16 = 256 RF entries. There are 32 such group to constitute 32-bits/RF entry by selecting only one cell/GBL. 

During read/write operation, the RS/WS are sent to the array in the first cycle. In the second cycle, 256 D1 footed-domino buffers per port are triggered by $\phi_{1}$ signal, buffered from the incoming 50\% duty-cycle core clock $Clk$ and drive the decoded RS (Fig. \ref{read_select}) and WS (Fig. \ref{write_select}) signals across the 32-bit array width. Fig. \ref{phase_signal} shows the different phase signal generation from $Clk$. 

Fig. \ref{read_ports_4} shows two read ports on each side of the
storage cell for symmetric loading and optimal cell write stability \cite{symmetric_loading}. Furthermore, balanced pass transistors are used on each side of the storage cell to enable single-ended full-swing write operation. The write data $D_{in}$, is locally inverted to get its complement.

Fig. \ref{Local_bitline_sceme} and \ref{global_bitline_scheme} shows the full-swing LBL and GBL scheme. LBL is precharged to $V_{dd}$ when $\phi_{1d}$ (a delayed version of $\phi_{1}$) is low. If one of the RS is asserted and the corresponding stored data is `1', LBL discharges when $\phi_{1d}$ is high. Note that bitcell write operation occurs when $\phi_{1d}$ is high. Two LBLs are merged via a static NAND gate leading to a dynamic 16-way NAND-NOR. Note that LBL phase is fully time borrowable enabling the RS/WS signals to arrive into the LBL evaluate phase. The GBL is a dynamic 8-way OR which merges the LBL 2-input static NAND outputs to deliver a 32-bit word per read port. GBL evaluation is done when $\phi_{2}$ (inverted signal of $Clk$) is high to avoid precharge races and crowbar currents at the phase boundary, the GBL domino is footed by the clock transistor. GBL phase is also fully time borrowable (LBL NAND outputs arrive into GLB circuit when $\phi_{2}$ is high).

\subsection{RF Vulnerabilities}
Two back to back transistors of RF cell that stores information is prone to noise. It gets worse due to technology and $V_{dd}$ scaling. If the noise is more than the static noise margin, it can flip the stored data. Therefore, adversary can leverage this to corrupt the stored data. The domino logic implemented for RF makes it even more prone to noise \cite{Domino_noise}. For example, LBL or GBL discharge can be prevented or they can be discharged intentionally during evaluation. This will lead to a read error. 


\section{RF Trojans}
In this section, we present BC, RP and LBL Trojans which can cause data corruption and read errors along with their overheads.

\begin{figure}[b]
        \centering
        \vspace{-6mm}
        \begin{subfigure}[b]{0.4236\linewidth}
                \centering
                \includegraphics[width=0.99\linewidth]{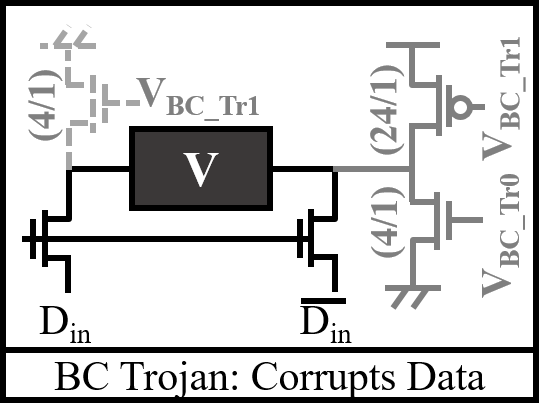}
                \caption{}
                \label{BC_trojan}
        \end{subfigure}%
        \hspace{-1mm}
        \begin{subfigure}[b]{0.556\linewidth}
                \centering
                \includegraphics[width=0.99\linewidth]{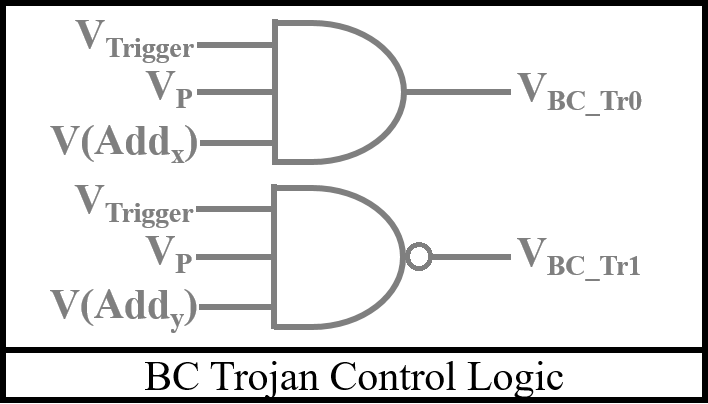}
                \caption{}
                \label{BC_trojan_control}
        \end{subfigure}\\
        \vspace{-2mm}
        \caption{BC Trojan: (a) payload circuit; (b) control logic.}
	\vspace{-2mm}
        \label{BC}
\end{figure}

\begin{figure}[t]
        \centering
        \vspace{-0mm}
        \begin{subfigure}[b]{0.49\linewidth}
                \centering
                \includegraphics[width=0.99\linewidth]{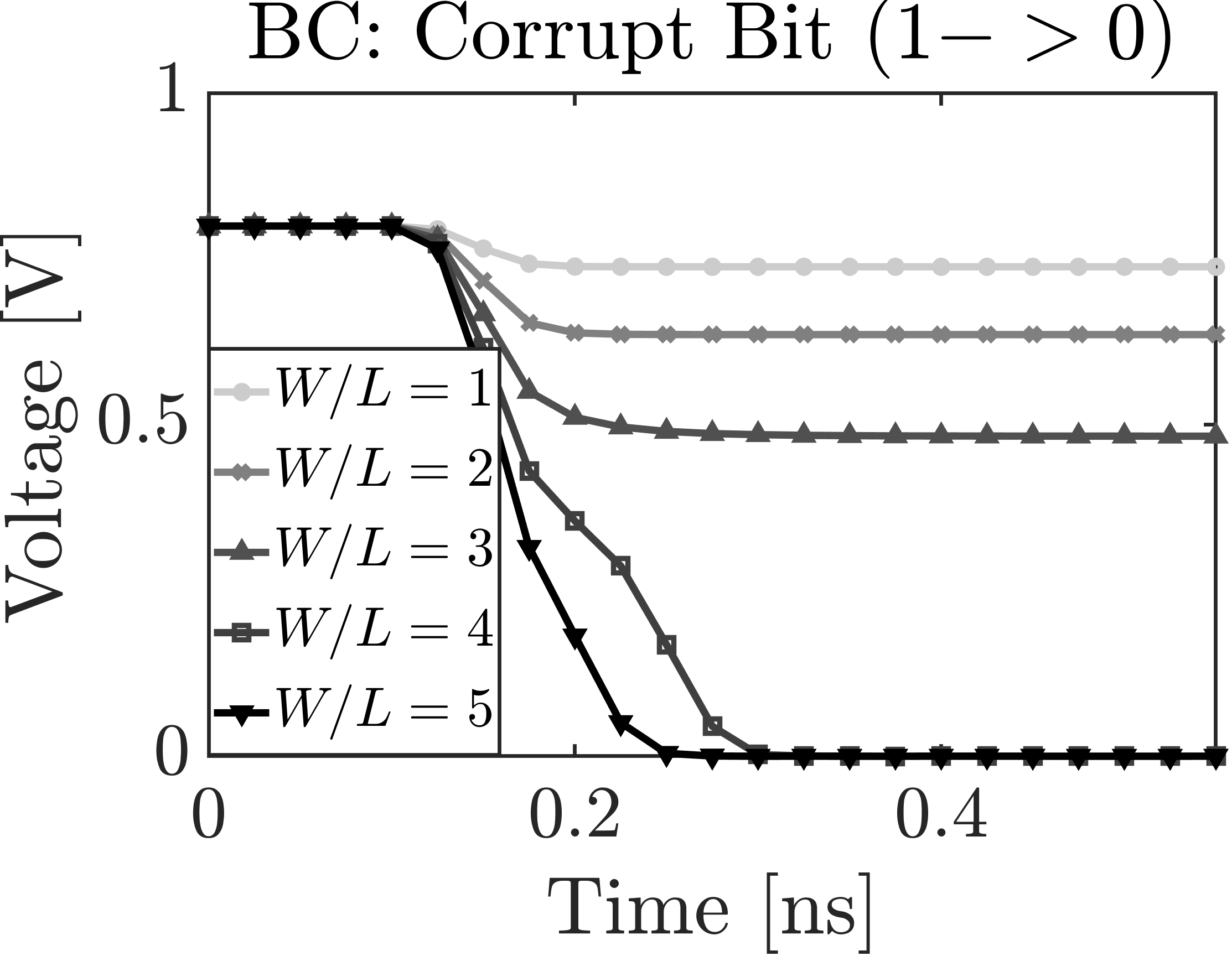}\vspace{-2mm}
                \caption{}
                \label{BC_1_to_0}
        \end{subfigure}%
        \hspace{-1mm}
        \begin{subfigure}[b]{0.49\linewidth}
                \centering
                \includegraphics[width=0.99\linewidth]{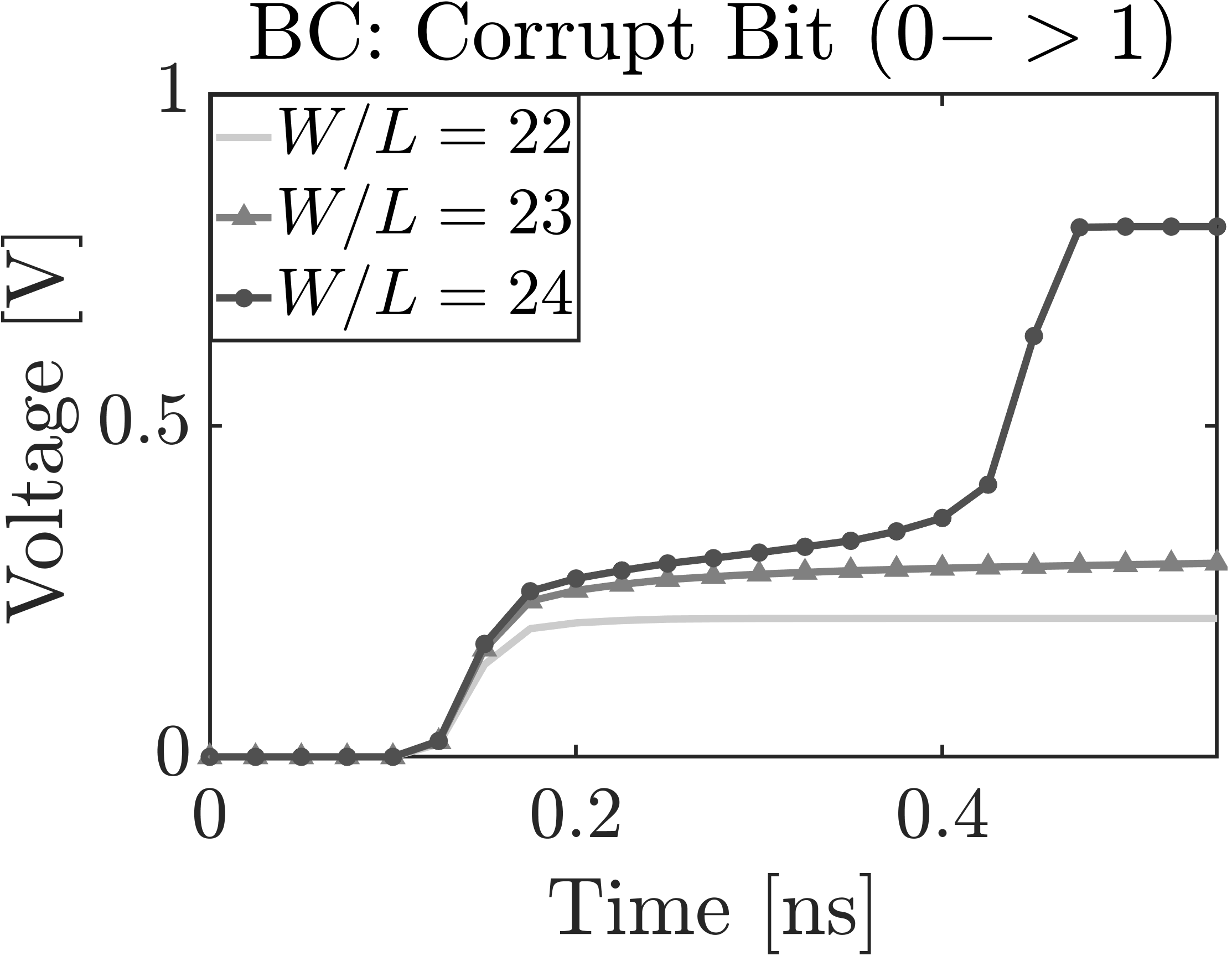}\vspace{-2mm}
                \caption{}
                \label{BC_0_to_1}
        \end{subfigure}\\
        \vspace{-3mm}
        \caption{(a) $1->0$ and (b) $0->1$ data resetting using BC Trojan.}
	\vspace{-2mm}
        \label{BC_results}
\end{figure}

\begin{figure} [t] 
\vspace{-0mm}
 \begin{center}
    \includegraphics[width=.49\textwidth]{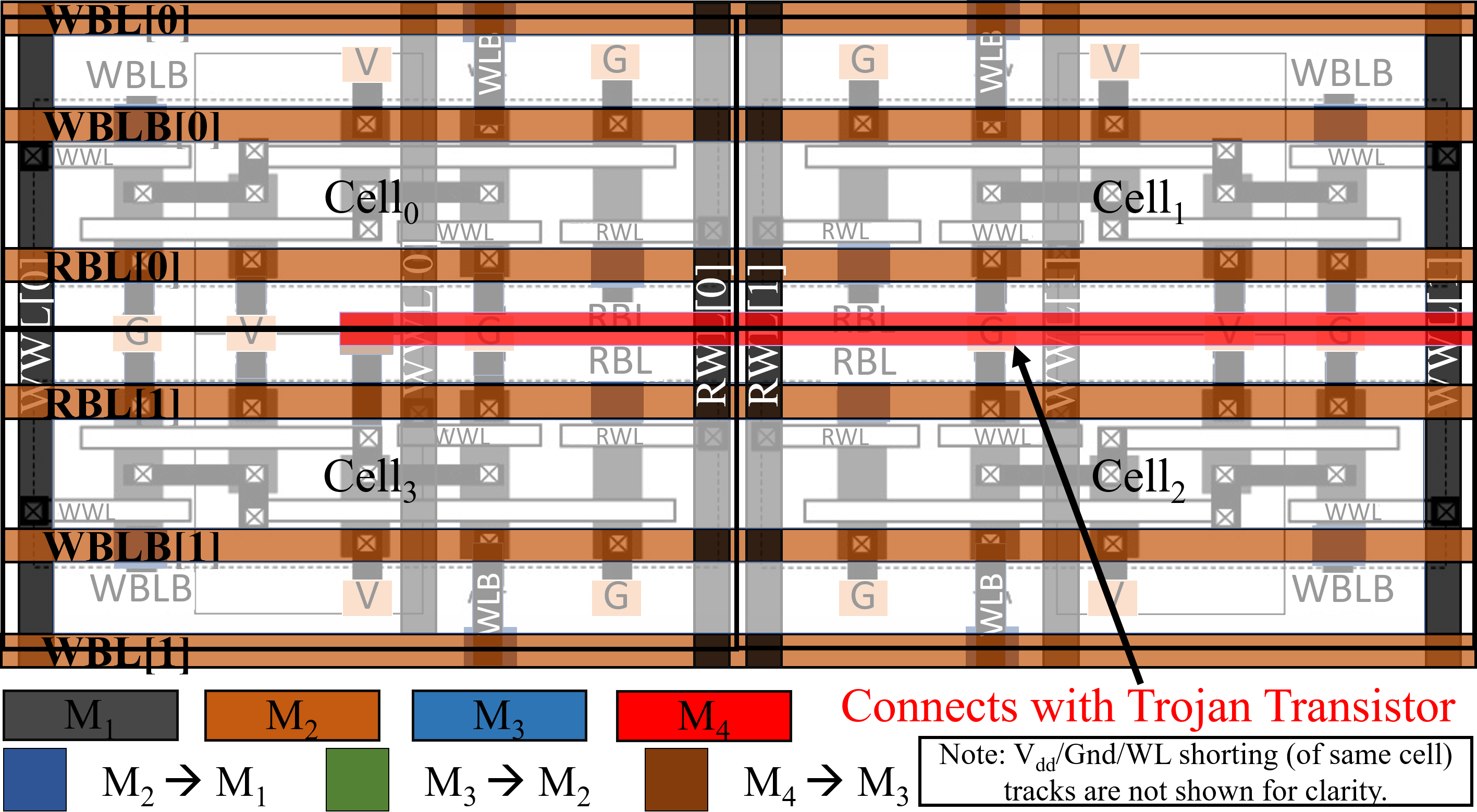}
 \end{center}
 \vspace{-3mm}
 \caption{BC Trojan: Layout showing 4 RF bitells (with one read/ write port) and metal tracks allocated for bitlines (horizontal) and wordlines (vertical). One $M_4$ track can be stolen to connect RF data node to Trojan payload transistor (located in the column area).} \label{layout}
 \vspace{-6mm}
\end{figure}

\subsection{BC Trojan}
RF bits can be reset to `0' or `1' when the cells are at retention. Fig. \ref{BC_trojan} shows the proposed BC Trojan (in grey color) that can corrupt the stored bit. Fig. \ref{BC_trojan_control} shows the logic to generate the control signal for BC Trojan. Once the Trojan trigger is activated (i.e., $V_{Trigger}$ = 1), writing $Add_x$ with $P_{SET}$ data pattern asserts $V_{BC\_Tr0}$ which can be leveraged to turn ON the NMOS and short the data node of RF to gnd. However, if $Add_y$ is written with $P_{SET}$ data pattern, $V_{BC\_Tr1}$ will be asserted which can be leveraged to short the data node to $V_{dd}$ via a PMOS. Fig. \ref{BC_1_to_0} and \ref{BC_0_to_1} show that the minimum (W/L) of NMOS and PMOS Trojan to reset the stored bit to `0' and `1' are 4 and 24 respectively. Alternative way to reset to data `1' is shorting the $\overline{data}$ node to gnd via an NMOS which incurs less area overhead (Trojan shown in Fig. \ref{BC_trojan} with dotted line). For a 32-bit word RF reset operation, 32 such transistor-pairs are required. However, if the reset pattern is preselected, only one transistor per bit is required. Fig. \ref{layout} shows the layout of 4 RF bitells (having one read/one write port) and metal tracks allocated for bitlines (horizontal) and wordlines (vertical). One $M_4$ track per global column  connects the target RF data node to the Trojan payload transistor which is co-located with sense-amp and other peripherals in the column area.

\begin{figure}[b]
        \centering
        \vspace{-6mm}
        \begin{subfigure}[b]{0.4646\linewidth}
                \centering
                \includegraphics[width=0.99\linewidth]{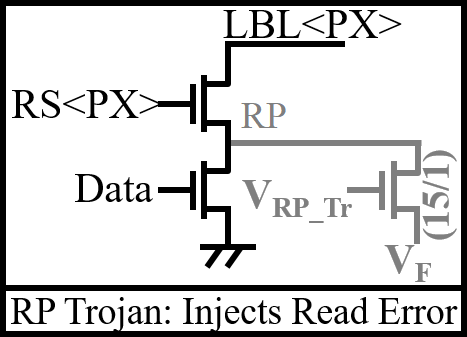}\vspace{-2mm}
                \caption{}
                \label{RP_trojan}
        \end{subfigure}%
        \hspace{-1mm}
        \begin{subfigure}[b]{0.5153\linewidth}
                \centering
                \includegraphics[width=0.99\linewidth]{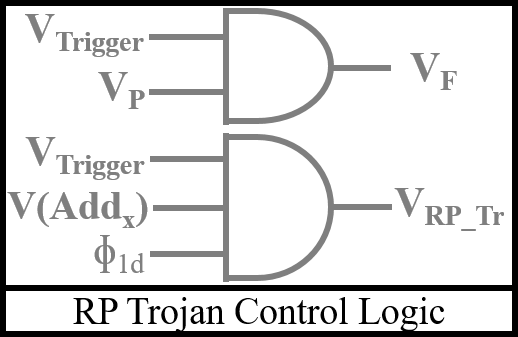}\vspace{-2mm}
                \caption{}
                \label{RP_trojan_control}
        \end{subfigure}\\
        \vspace{-3mm}
        \caption{RP Trojan: (a) payload circuit; (b) control logic.}
	\vspace{-2mm}
        \label{RP}
\end{figure}

\begin{figure}[t]
        \centering
        \hspace{0mm}
        \begin{subfigure}[b]{0.49\linewidth}
                \centering
                \includegraphics[width=0.99\linewidth]{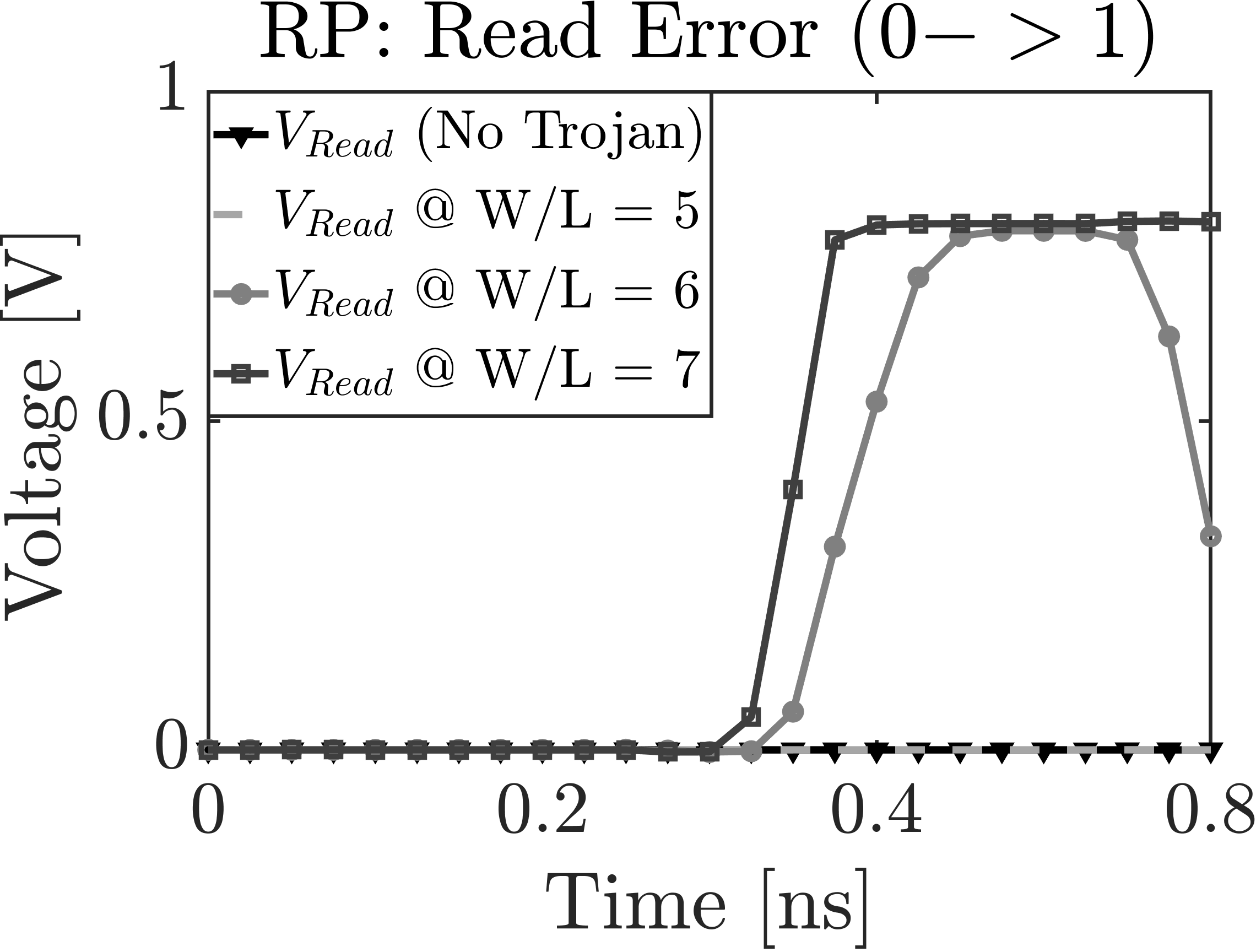}\vspace{-2mm}
                \caption{}
                \label{RP_0_to_1}
        \end{subfigure}%
        \hspace{-1mm}
        \begin{subfigure}[b]{0.49\linewidth}
                \centering
                \includegraphics[width=0.99\linewidth]{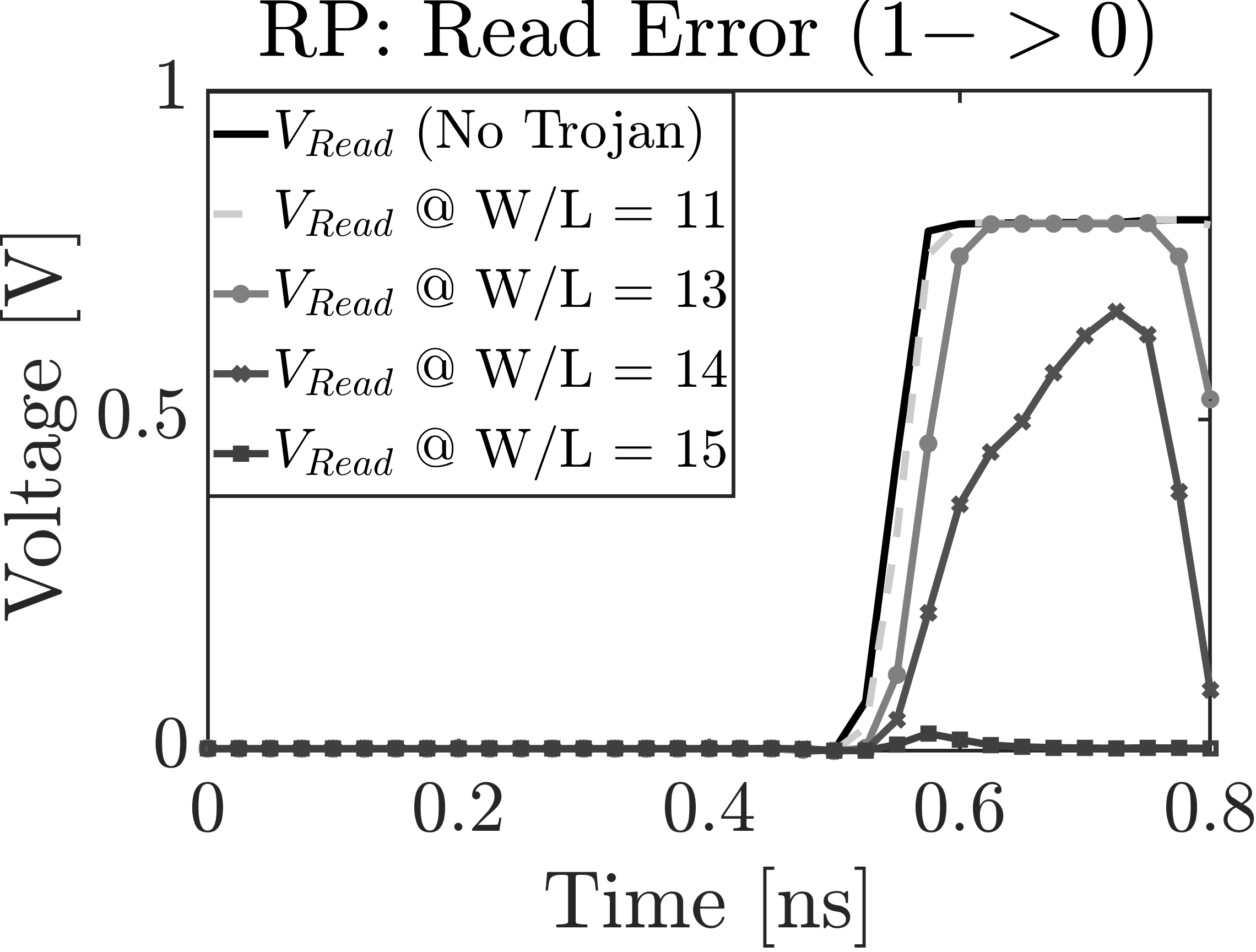}\vspace{-2mm}
                \caption{}
                \label{RP_1_to_0}
        \end{subfigure}\\
        \vspace{-3mm}
        \caption{(a) $0->1$ and (b) $1->0$ read error using RP Trojan.}
	\vspace{-2mm}
        \label{RP_results}
\end{figure}

\begin{figure} [t] 
\vspace{-0mm}
 \begin{center}
\includegraphics[width=.49\textwidth]{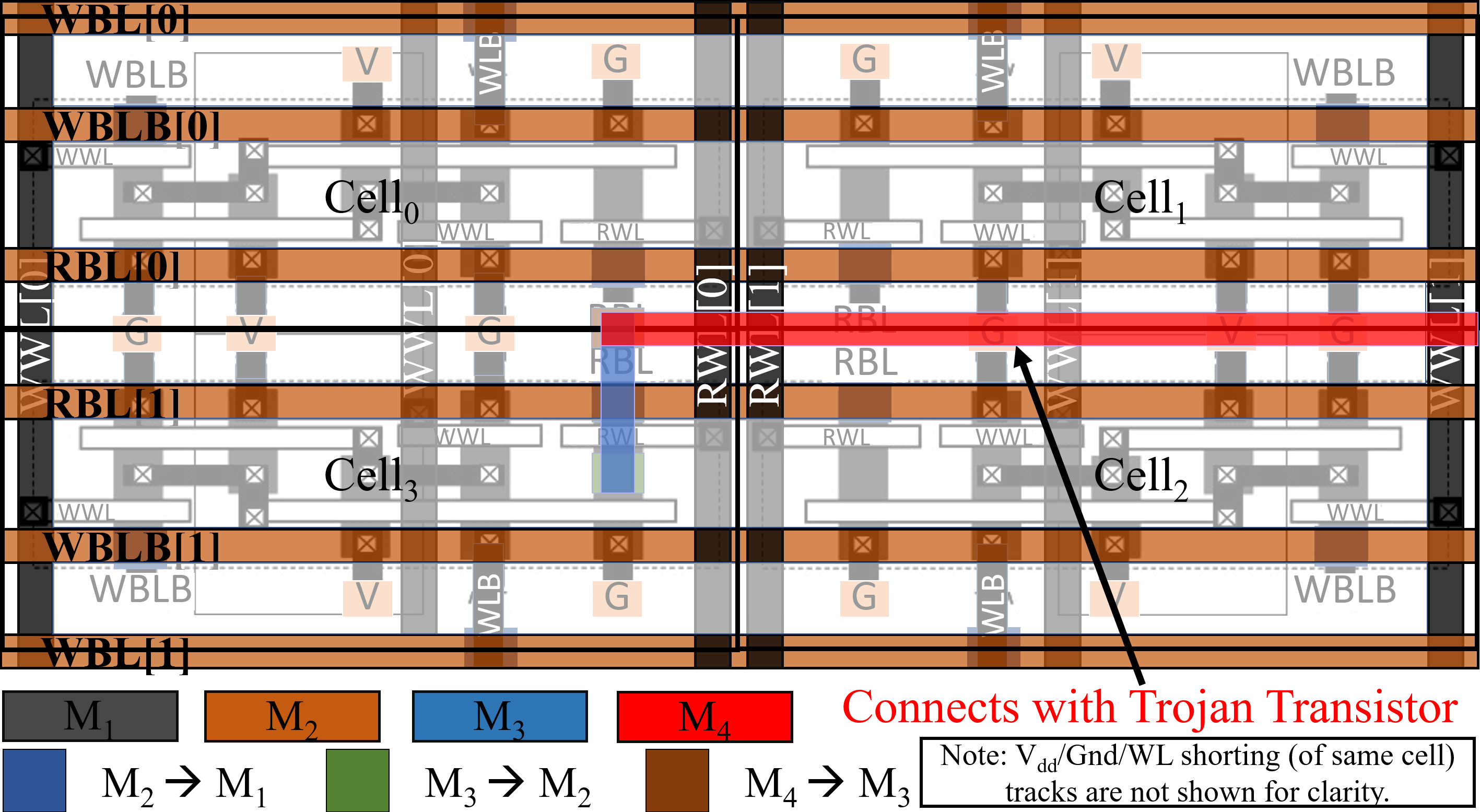}
 \end{center}
 \vspace{-4mm}
 \caption{RP Trojan: one extra $M_3$ track is needed compared to BC.} \label{layout_RP}
 \vspace{-6mm}
\end{figure}

\subsection{RP Trojan}
RP Trojan (Fig. \ref{RP_trojan}) can be implemented if adversary requires error injection during read operation. In this case, only one NMOS Trojan is required to inject both type of read errors. $V_{RP\_Tr}$ can be asserted if $Add_{x}$ is written after Trojan trigger is activated (i.e. $V_{Trigger} = 1$) during LBL evaluation period ($\phi_{1d}$ = 1) (Fig. \ref{RP_trojan_control}). If the write data pattern is $P_{SET}$, $1->0$ read error will be injected (since $V_F$ = 1). Otherwise, $V_F$ will be 0 and $0->1$ read error will be injected. Fig. \ref{RP_0_to_1} and \ref{RP_1_to_0} show that the minimum (W/L) of NMOS Trojan to inject $0->1$ and $1->0$ read error are 6 and 15 respectively. If a single Trojan is implemented for both type of errors, the minimum (W/L) should be 15. Fig. \ref{layout_RP} shows the layout to insert RP Trojan. In this case, we need one additional $M_3$ track compared to BC Trojan. Note that the same bit can be read through different port differently with RP Trojan.

\begin{figure}[b]
        \centering
        \vspace{-5mm}
        \begin{subfigure}[b]{0.4941\linewidth}
                \centering
                \includegraphics[width=0.99\linewidth]{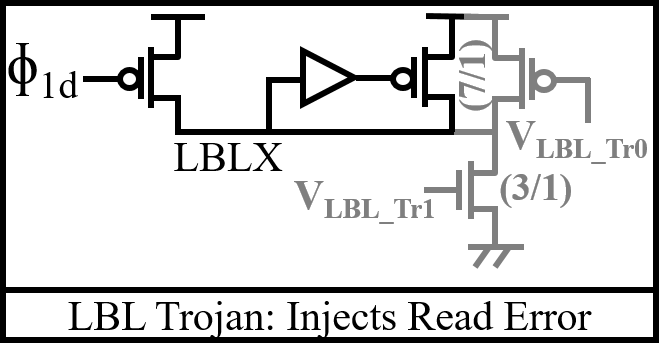}
                \caption{}
                \label{LBL_trojan}
        \end{subfigure}%
        \hspace{-1mm}
        \begin{subfigure}[b]{0.4859\linewidth}
                \centering
                \includegraphics[width=0.99\linewidth]{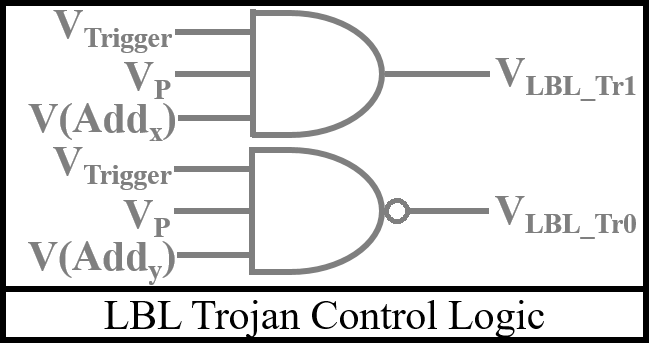}
                \caption{}
                \label{LBL_trojan_control}
        \end{subfigure}\\
        \vspace{-2mm}
        \caption{LBL Trojan: (a) payload circuit; (b) control logic.}
	\vspace{-4mm}
        \label{LBL}
\end{figure}

\begin{figure}[b]
        \centering
        \hspace{-1mm}
        \begin{subfigure}[b]{0.49\linewidth}
                \centering
                \includegraphics[width=0.99\linewidth]{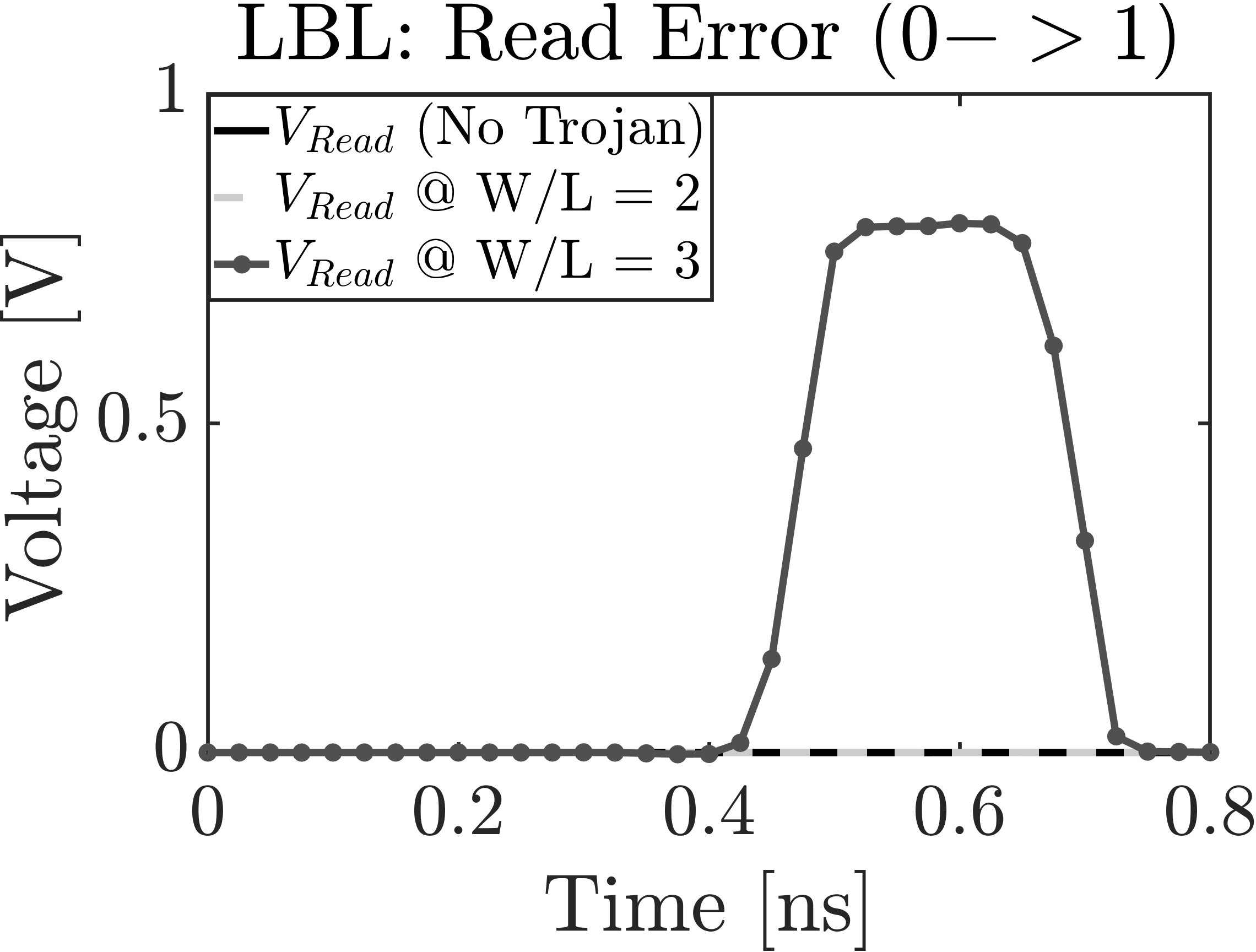}
                \caption{}
                \label{LBL_0_to_1}
        \end{subfigure}%
        \hspace{-1mm}
        \begin{subfigure}[b]{0.49\linewidth}
                \centering
                \includegraphics[width=0.99\linewidth]{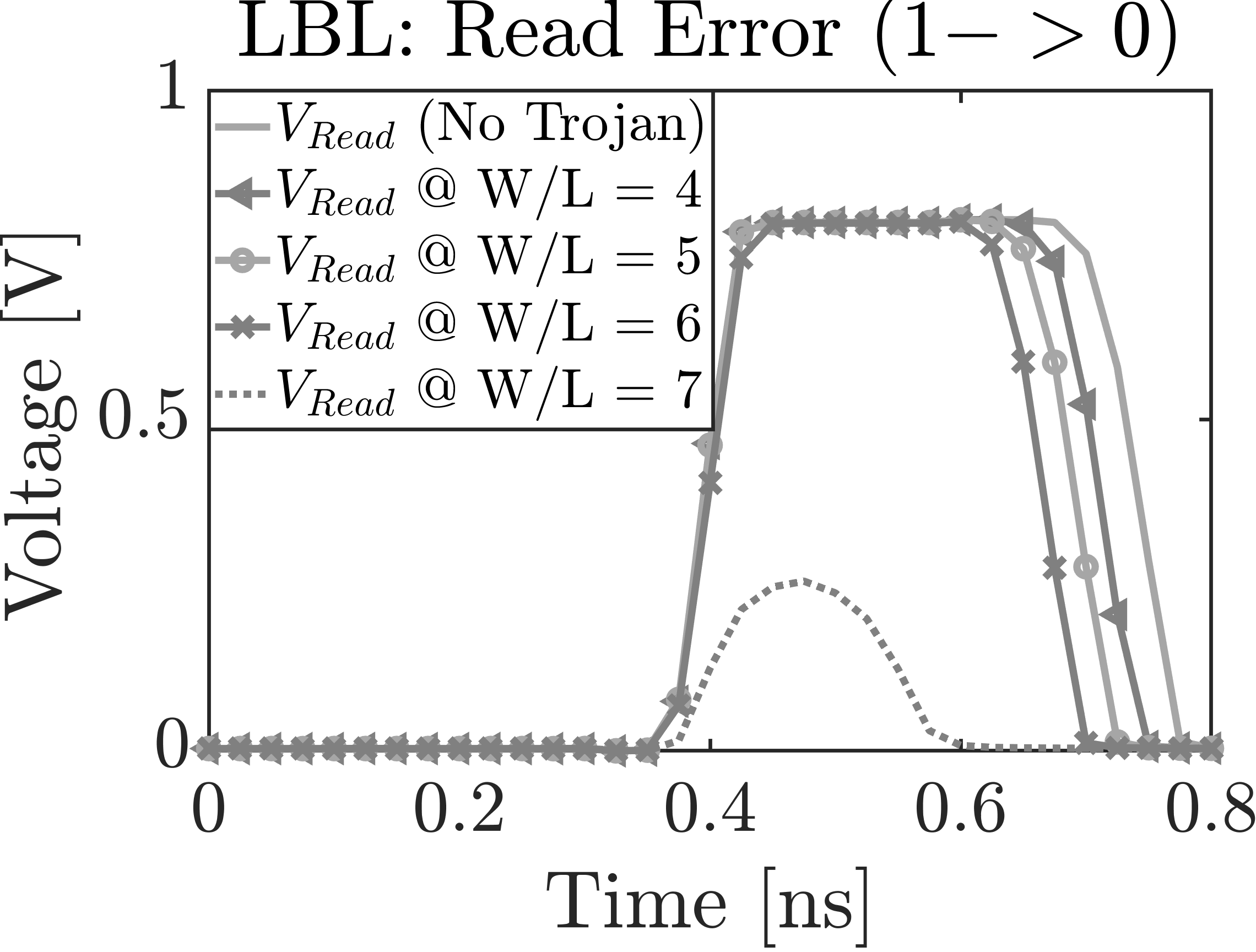}
                \caption{}
                \label{LBL_1_to_0}
        \end{subfigure}\\
        \vspace{-2mm}
        \caption{(a) $0->1$ and (b) $1->0$ read error using LBL Trojan.}
	\vspace{-2mm}
        \label{LBL_results}
\end{figure}

\subsection{LBL Trojan}
Faults can be injected during read operation of RF, i.e. data `0' will be read as `1' (or vice-versa) using the proposed LBL Trojan (Fig. \ref{LBL_trojan}). NMOS and PMOS Trojans are required to inject read $1->0$ and $0->1$ faults respectively which can be turned ON using $V_{LBL\_Tr0}$ and $V_{LBL\_Tr1}$ (Fig. \ref{LBL_trojan_control}) respectively. Similar to RP Trojan, $V_{LBL\_Tr0}$ and $V_{LBL\_Tr1}$ will only be asserted when $\phi_{1d}$ is high during the read operation (LBL evaluation period). Fig. \ref{LBL_0_to_1} and \ref{LBL_1_to_0} show that the minimum (W/L) of NMOS and PMOS Trojan to inject $0->1$ and $1->0$ read errors are 3 and 7 respectively. Note that unlike RP Trojan, in LBL Trojan, all the 16 bits (each of different RF entry, Fig. {\ref{Local_bitline_sceme}}) connected with the Trojan-infected LBL will incur the read error if those cells are read when $V_{LBL\_Tr1}$ = 1 or $V_{LBL\_Tr0}$ = 0 (payload deployed). Furthermore, the Trojan can be implemented in the peripheral area where the component density is slightly relaxed.

\subsection{Overhead Analysis}
Table \ref{Trojan_overhead} summarizes the area/power overheads for 2-bit Trojans (to corrupt the CPL field) along with the required control logic. However, we didn't consider the parasitics since they only affect the delay. Note that all the Trojans can be implemented in the peripheral area where there are enough empty space. For BC Trojan, one $M_4$ and for RP Trojan, one $M_3$ and one $M_4$ track needs to be routed to column area. We can conclude that the static power overheads of the Trojans are insignificant. The dynamic power of some cases is high. However, they will only occur one time when they are activated. Trojan use cases (Table {\ref{Trojan_overhead}}) are explained in Section VI.

\begin{table}[t]
\caption{Features of the Proposed 2-bit RF Trojans}
\vspace{-4mm}
\begin{center}
\begin{tabular}{|c|c|c|c|c|}
\hline
	\textbf{Trojan} & \textbf{Static} & \textbf{Dynamic} & \textbf{Area} & \textbf{Use}\\ 
	 &  \textbf{Power (nW)} &  \textbf{Power ($\mu$W)} &  \textbf{($\mu$m\textsuperscript{2})} & \textbf{Case}\\
	\hline\hline
	\textbf{BC} (\textbf{$0\rightarrow1$}) & 0.079 & 62.44 & 0.056 & Kernel \\ 
	\textbf{BC} (\textbf{$1\rightarrow0$}) & 0.083 & 8.37 & 0.023 & Leak\\ \hline
	\textbf{RP} (\textbf{$0\rightarrow1$}) & 12.93 & 45.38 & 0.048 & Kernel  \\ 
	\textbf{RP} (\textbf{$1\rightarrow0$}) & 33.73 & 112.34 & 0.064 & Leak\\ \hline
	\textbf{LBL} (\textbf{$0\rightarrow1$}) & 35.28 & 57.54 & 0.022 & \multirow{2}{*}{DoS} \\
	\textbf{LBL} (\textbf{$1\rightarrow0$}) & 11.26 & 24.57 & 0.026 & \\ \hline

\end{tabular}
\vspace{-3mm}
\label{Trojan_overhead}
\end{center}
\vspace{-4mm}
\end{table}

\begin{figure}[b]
\vspace{-2mm}
 \begin{center}
    \includegraphics[width=0.49\textwidth]{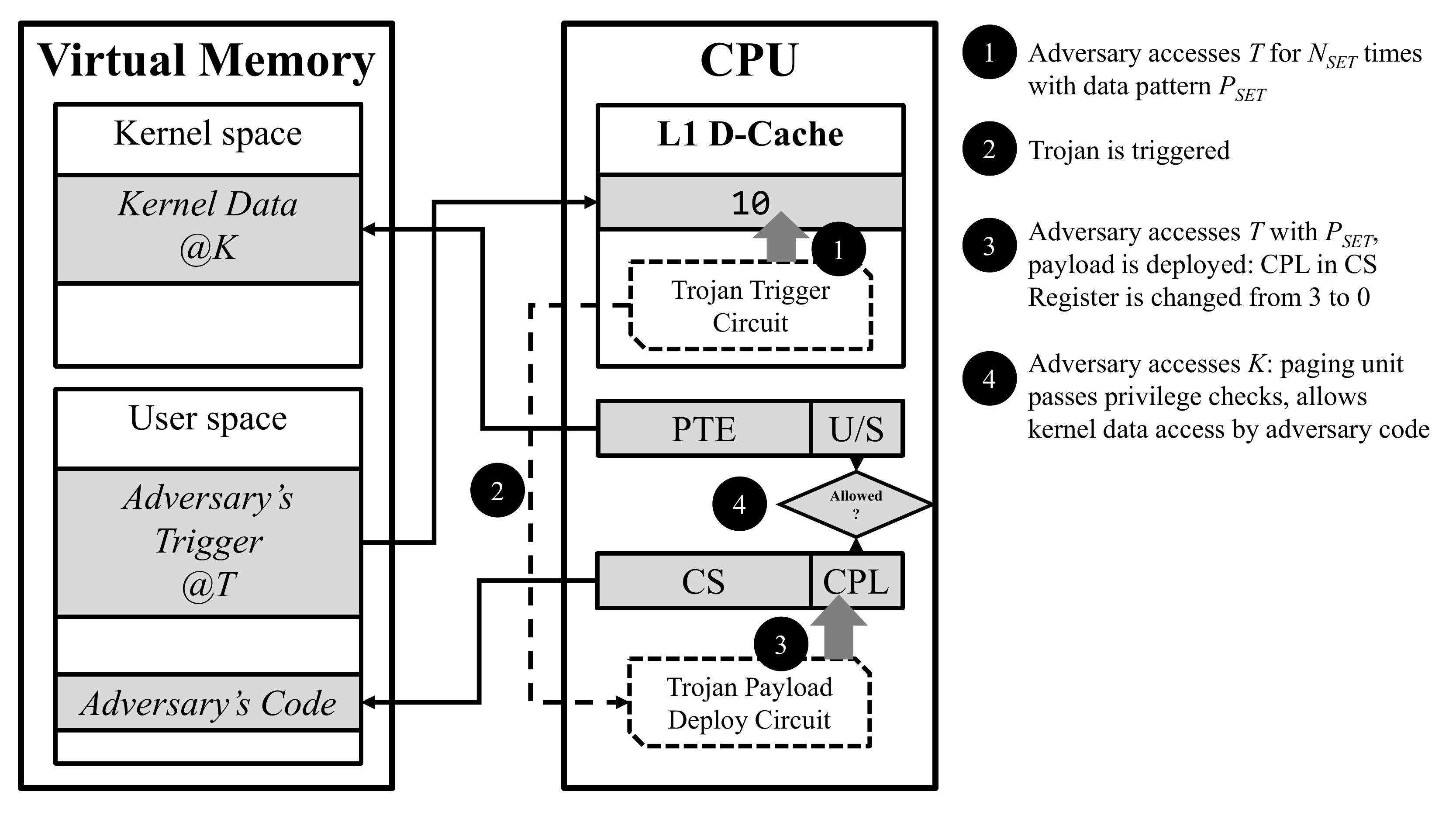}
 \end{center}
 \caption{Trojan design in the system architecture with attack steps.} \label{trojatk}
\end{figure}

\begin{figure}
\vspace{-3mm}
\begin{center}
    \includegraphics[trim=0 0 7.7in 0,clip,width=0.49\textwidth]{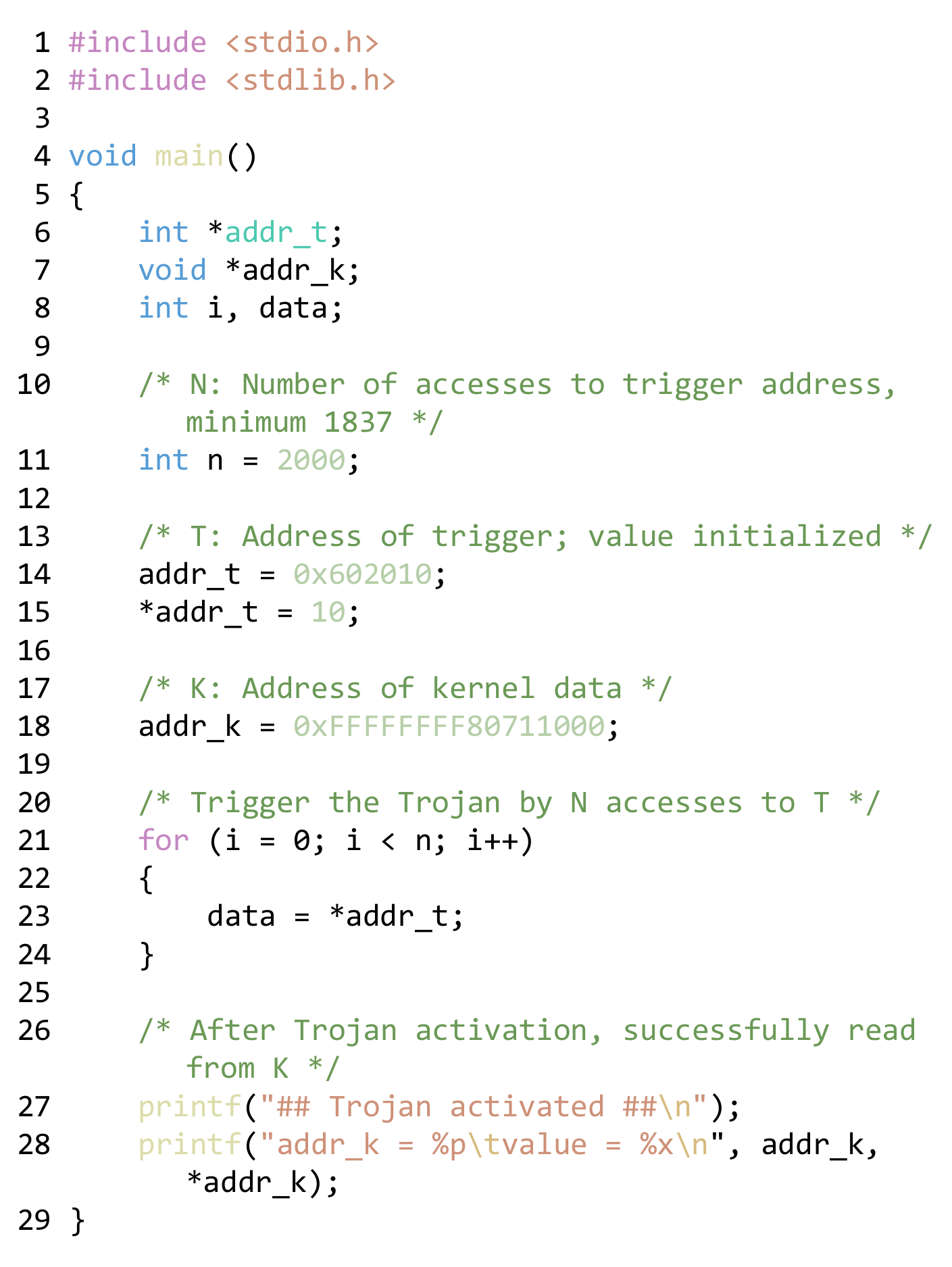}
 \end{center}
\vspace{-2mm}
\caption{Example exploit code using the RF-Trojan.}
\label{attackcode}
\end{figure}

\section{Privilege Escalation using RF-Trojan}
We describe a simple proof-of-concept attack demonstrating the use of the BC Trojan. The goal of the attack is to gain access to and read data from kernel data pages. The malicious RF-Trojan is deployed in the system architecture as shown in Fig. \ref{trojatk}. With the knowledge of the RF-Trojan, the adversary runs his exploit code. The adversary's code runs in user mode and the CPL in the CS register is set to \texttt{b'11} (3) accordingly. The adversary process initially accesses a `trigger' address $T$ ($Add_{SET}$) for at least $N_{SET}$ times with $P_{SET}$ pattern (here, \texttt{10}). The trigger is placed in a L1 d-cache line (which maps to $T$), that monitors its access for $N_{SET}$ times and activates the trigger (i.e. $V_{Trigger}$ = 1). Next, an additional access to $T$ ($Add_{X}$) with $P_{SET}$ pattern generates a signal for the Trojan payload deployment (i.e. $V_{BC\_Tr0}$ = 1). For simplicity, we have used the address $T$ for both $Add_{SET}$ and $Add_{X}$. The Trojan payload is placed in the CS register in the CPU. After the payload is deployed, CPL field is set to \texttt{b'00} (0), which maliciously elevates the privilege of the adversary's code to ring 0. The adversary can now successfully access and read data from an address $K$ in the kernel pages (with U/S bit set to 1), since the paging unit now thinks that the CPU is in CPL 0. The exploit code is shown in Fig. \ref{attackcode}. Without deploying the payload (lines 21-24), reading from \texttt{addr\_k} in line 28 will cause a segmentation fault, since the corresponding kernel page has the U/S bit set to 1 in the PTE while the CPL for the code is 3.

\begin{figure} [t] 
\vspace{-0mm}
 \begin{center}
    \includegraphics[trim=0 5.75in 0 0,clip,width=0.49\textwidth]{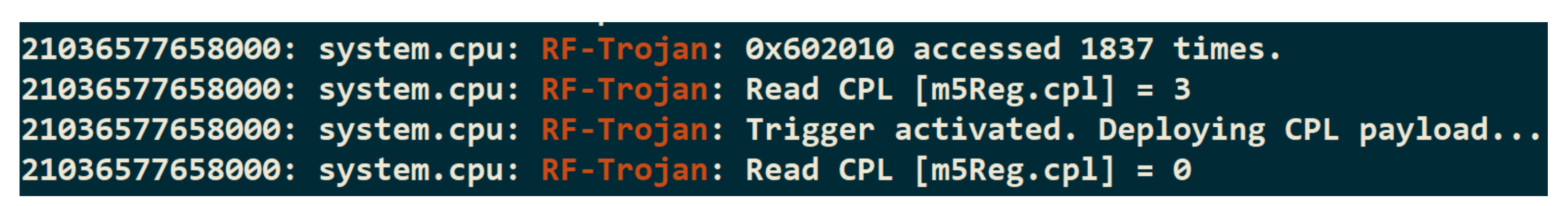}
 \end{center}
 \vspace{-4mm}
 \caption{GEM5 debug logs showing privilege escalation.} \label{debuglog}
 \vspace{-6mm}
\end{figure}

We model the RF-Trojan in GEM5 simulator in x86 architecture. We have considered an AtomicSimpleCPU model for the Trojan deployment. To simulate the Trojan trigger, we added code to the \texttt{readMem()} function in the CPU code to monitor accesses to trigger address \texttt{addr\_t} (\texttt{0x602010}). Since the L1 cache is virtually indexed, we can just monitor the cache line mapped to the trigger address. For the Trojan payload insertion, we added code to retrieve the current \texttt{m5} register value which maintains the CPL. We set the CPL field (\texttt{m5Reg.cpl}) to 0 and write back the value to the \texttt{m5} register. We added debug logs in the GEM5 code to verify the privilege escalation (Fig. \ref{debuglog}). After the privilege escalation of the adversary's process, reads from the kernel pages are made possible.

\section{Discussion}
\subsection{Other Possible Attack Models}
In this work, we limit our attack demonstration using BC Trojan for the sake of brevity. Attack model for other Trojans are given below considering Trojan trigger is already activated:

\textbf{Using RP Trojan:} Adversary accesses $Add_{x}$ with data pattern $P_{SET}$. Therefore, $V_{RP\_Tr0}$ remains 1 for a few clock cycles (equal to L1 cache latency). If CPL field is read during this time, paging unit will assume the CPU to be in kernel mode. The adversary writes his exploit in a way that after the trigger is activated, the process forks itself into multiple child processes, each of which try to gain access to kernel page addresses ($K$). Since RP Trojan injects error through only one read port, some of these processes will terminate (since they were given access through a non-tampered read port) while one of them gets access to address $K$ (which was given access through the Trojan injected read port). This is possible since RF latency is significantly lower than L1 cache latency and multiple RF reads are done while $V_{RP\_Tr0}$ is asserted high.

\textbf{Using LBL Trojan:} The attack is similar to RP Trojan attack except, adversary accesses $Add_{y}$ with $P_{SET}$ data pattern instead of $Add_{x}$ which makes $V_{LBL\_Tr0}$ = 0 for a few clock cycles (equal to L1 cache latency). Note that this Trojan injects error to all the other 15 RF bits along with the target bit that share the same LBL. If those bits are also read while $V_{LBL\_Tr0}$ = 0 which is highly likely (e.g., the index field in the segment registers which point to the current segment in use), the corresponding processes will crash. Therefore, LBL Trojan is more suitable for Denial of Service (DoS) attacks rather than kernel data leakage.


\textbf{Exploiting General Purpose Registers (GPRs):} GPRs can also be targeted for deploying the Trojan payload. Assuming a victim process that accesses the trigger address where the number of accesses is controlled by user input, the adversary can trigger the Trojan through the input. If the payload is deployed to a GPR (e.g., eax, ebx, etc.), the adversary can force an erroneous computation by an instruction that uses that GPR. In case of a register indirect addressing that uses the victim register, the adversary can force the address pointer to point to a specific address (deployed as the Trojan payload) in the victim process's address space and leak sensitive data from that address. Deploying payloads to the registers for instruction pointer (eip), stack (esp) and base (ebp) pointers may also be used to cause control-flow violations.

\subsection{BC Trojan Vs RP Trojan}
Attack using BC Trojan is permanent. Once the attack is launched, the CPL field will remain modified until there is a context switch, when the register is re-written and the Trojan trigger is de-asserted. However, attack using RP can be sneaky and transient since the attack can launch, get access to kernel data and come back to normal operation. Also, the content of RF remains correct. However, RP Trojan overheads is higher compared to BC. Therefore, adversary can implement either one based on the use case and design constraints to leak kernel data.

\subsection{Countermeasures}

\textbf{Read verification leveraging multi-port feature:} RF inherently has the multiple read ports. We propose to use one port to read a RF entry after it is read by another port. This will capture any read failure caused by RP/LBL Trojans and raise an exception. This can be done in two ways: 

i) Implement a dedicated verification port: This will incur area/power overhead. However, since the verification can be done concurrently, this will not hurt the system throughput and detect any failure caused by hardware Trojan. Furthermore, if the Trojan payload is deployed to CPL only, it is likely that the CS register is read through a single port, since, (i) only the kernel can read CPL value, and  (ii) CPL validation is performed in paging unit. Hence, CPL verification only needs to be done with a single port.

ii) Opportunistically assign a port for verification: While accessing one RF entry, another unused read port can be dynamically assigned for verification. This will incur some performance overhead since the port assigned for verification will be unavailable for other operations until it is freed. Furthermore, if all the read ports are already busy, none can be assigned for verification and attack can remain undetected. However, this saves design overhead.

\textbf{Securing control and segment registers:} A Physically Unclonable Function (PUF)-based secure hash can be implemented which will store a hash of RF write data in a secure memory. An arbiter or memory PUF can be used. Note that compared to an actual hash function, the proposed PUF-based hash can be area and performance efficient. The CR address acts as a challenge and the response acts as a hash. Whenever this data is read, a hash of read data can be compared with the previously stored hash of write data. This detects any read (by RP/LBL Trojans) or retention (by BC Trojan) failures indicating an attack. Note that we propose this only for the control (e.g., CR0-CR4 in x86) and segment registers (e.g., CS, SS, DS, etc. in x86) and not for the GPRs for two reasons: i) less overhead since PUF can be small to cater to the limited number of narrow width control and segment registers (32-bit and 16-bit wide respectively) compared to large number of 64-bit wide GPRs; ii) control/segment registers determine the CPU operation mode and ensuring their security is far more critical than GPRs.

\textbf{L1 address obfuscation:} Typically, L1 cache is virtually indexed and physically tagged. However, this is a vulnerability since adversary can hammer L1 cache using virtual address. Therefore, L1 address obfuscation (using a PUF, for example) to change virtual to physical mapping of L1 cache can add a layer of complexity on the adversary. This is true since the predefined memory address can no longer be hammered. Similar technique has been shown to be effective in preventing side-channels in the cache \cite{newcache}.

\subsection{Limitations and Possible Issues}

\textbf{Context switching:} GPRs are very frequently used and hence it might be limited as a target to launch real exploits. This is because, the values of GPRs are backed up and flushed in case of a context-switch. Hence, changing a GPR while the adversary's attack code is running, only affects the adversary's code and not a victim process's code. The alternate strategy is to wait until the victim process comes into context and then inject the fault in the GPR. However, this will still be more effective in causing DoS attacks than data leakage. 

\textbf{Register allocation:} Out-of-order pipelines have a register allocation unit for GPRs which maps architectural registers to physical registers based on availability. Hence, it might be difficult for the adversary to control the access to the physical register. The adversary can perform read/write on only the architectural registers specified by the ISA, and thus cannot guarantee that the Trojan deployed physical register will be used to make his attack successful.

\section{Conclusions}
In this work, we investigated the RF vulnerabilities that can be exploited to implement Trojans. We proposed  BC Trojan to cause retention failure and RP/LBL Trojans to inject read error. We demonstrate a privilege escalation attack using an exploit code that uses the RF-Trojan to modify a critical segment register. Finally, we propose techniques to detect/prevent such Trojans.

\addtolength{\textheight}{-12cm}   








\bibliographystyle{ieeetr}
\bibliography{sample-bibliography}

\end{document}